\documentclass[amsmath, amssymb, aps, prfluids,
superscriptaddress, onecolumn, longbibliography, 10pt]{revtex4-2}

\usepackage{graphicx}%
\usepackage{amsmath}%
\usepackage{xcolor}%
\usepackage{bm}
\usepackage{hyperref}

\newcommand{\pdiff}[2]{\ensuremath{\frac{\partial {#1}}{\partial {#2}}}}

\newcommand{\vn}{\ensuremath{v_\mathrm{n}}}
\newcommand{\vs}{\ensuremath{v_\mathrm{s}}}

\newcommand{\bv}[1]{\ensuremath{\mathbf{#1}}}

\newcommand{\rhos}{\ensuremath{\rho_s}}
\newcommand{\rhon}{\ensuremath{\rho_n}}

\begin{document}

\title{Critical behavior and multistability in quasi-two-dimensional turbulence}

\author{Filip Novotn\'{y}}
\author{Marek Tal\'{i}\v{r}}
\author{\v{S}imon Midlik}
\author{Emil Varga}
\email{emil.varga@matfyz.cuni.cz}
\affiliation{Faculty of Mathematics and Physics, Charles University, Ke Karlovu 3, Prague, 121 16, Czech Republic}

\begin{abstract}
    Two-dimensional (2D) turbulence, despite being an idealization of real flows, is of fundamental interest as a model of the spontaneous emergence of order from chaotic flows. The emergence of order often displays critical behavior with the involvement of a broad range of spatial and temporal scales. Here, we experimentally study turbulence in periodically driven nanofluidic channels with a high aspect ratio using superfluid helium. We find a multistable transition behavior resulting from cascading bifurcations of large-scale vorticity and critical behavior at the transition to quasi-2D turbulence consistent with phase transitions in periodically driven many-body systems. We demonstrate that quasi-2D turbulent systems can undergo an abrupt change in response to a small change in a control parameter, consistent with predictions for large-scale atmospheric or oceanic flows.
\end{abstract}

\maketitle

\section{Introduction}
\label{sec:main-intro}

Chaotic, turbulent flows are ubiquitous and occur in vastly diverse systems ranging from a flow around a swimming penguin \cite{Masud2022} to climate systems \cite{Boffetta2012,Pedlosky2013}. In homogeneous isotropic turbulence in three dimensions, energy is transported to small scales until dissipated by viscosity, thus destroying any large-scale features \cite{Davidson2015}. However, in two dimensions, turbulence transports energy to large scales in an inverse cascade, which can result in the emergence of large-scale coherent structures called condensates in analogy with Bose-Einstein condensation \cite{Boffetta2012}. The condensate state is, generally, not unique, and stochastic changes between distinct metastable states, such as zonal jets or large-scale vortices, are possible \cite{Bouchet2009}. These transitions are rare \cite{Bouchet2019}; however, recent extensive simulations \cite{Xu2024} show a transition between preferred states as Reynolds number and flow aspect ratio is varied with fluctuating transitions in the intermediate region. Allowing for finite thickness of the flow, the possible phase transition-like phenomena in turbulent flows become very rich \cite{vanKan2024}. However, even the existence of a critical point for the lifetime or the formation time of the vorticity condensate is uncertain, where scenarios with zero, one or two critical points are possible \cite{Kan2019}.

Experimentally, phase transitions were observed in 3D turbulent flows \cite{Cortet2010}, and multistability of turbulent flows was observed as spontaneous reversals in magnetically forced turbulence \cite{Sommeria1986,Michel2016} in thick layers, and for flows in the stratosphere \cite{Schmeits2001} and their laboratory models \cite{Semin2018}. In thin layer flows, experiments focused mostly on the build-up and long-time behavior of vorticity condensates and their influence on velocity statistics in the inverse cascade \cite{Xia2009,Xia2017}. Long-lived bistability was observed in oscillatory flows \cite{Varga2020}, although no stochastic transitions were present.

The experimental study of intermittent transitions is often limited by the aspect ratios required for two-dimensionality and the rarity of the transitions. In this work, we aim to shed light on the critical behavior of quasi-two-dimensional turbulence, which is thus far accessible only in numerical simulations. Our experimental method allows long observation times compared to typical timescales of the turbulent flow. We observe the stochastic transitions between bistable states and strong multistability involving at least 8 turbulent states in driven oscillatory turbulent flow of superfluid helium confined to channels of aspect ratio $2000D\times 2000D \times D$, $D\approx 500$~nm. The lifetimes of the turbulent states scale with the flow velocity in the bistable regime approximately as a power-law with critical exponent consistent with random organization \cite{Corte2008} or 2D directed percolation \cite{Hof2023}, which supports the scenarios with at least one critical point for the formation of the condensate suggested in \cite{Kan2019}.

The paper is organized as follows: In section \ref{sec:main-setup}, the Helmholtz resonators and the measurement of vortex line density are briefly reviewed; in section \ref{sec:main-results-vc1}, we discuss the first critical velocity for the onset of nonlinear damping, which is likely connected with depinning of quantized vortices from the surface; in section \ref{sec:main-results-multistability} we show the growth of vortex line density with superflow velocity in the transitional region and the appearance of multistability depending on the geometry and temperature and develop a dynamical model that qualitatively captures the observed behavior; in section \ref{sec:main-results-transitions} the stochastic transitions between two different flow states are shown whose lifetimes approximately follow diverging power laws after which the discussion and conclusions follow.

\section{Experimental Setup}
\label{sec:main-setup}

\begin{figure}
    \centering
    \includegraphics{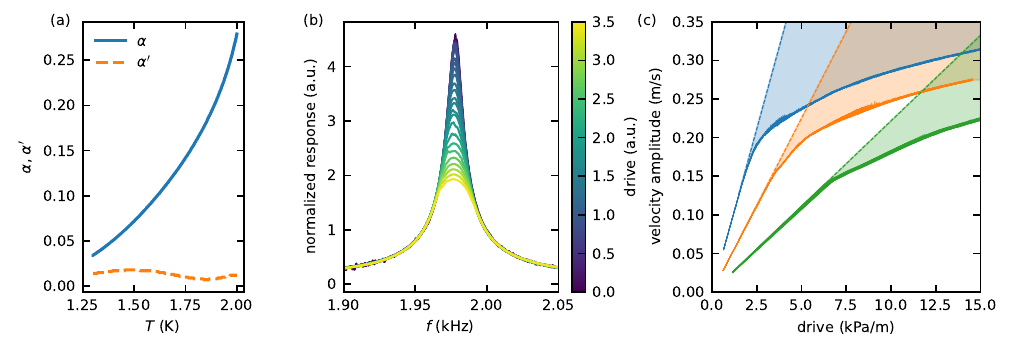}
    \caption{Nonlinear response of the Helmholtz resonators. (a) Mutual friction coefficients entering \eqref{eq:vs-flow} from ref. \cite{Donnelly1998}. (b) Typical resonant response of the device normalized by the drive. Suppression of the peak indicates nonlinear dissipation. (c) Peak velocity as a function of the driving amplitude of device \emph{C} at 1.45~K, 1.65~K, and 1.85~K (top to bottom); shaded areas indicated the excess nonlinear damping.}
    \label{fig:response}
\end{figure}

We use superfluid $^4$He (He-II) as the working medium, whose lack of viscosity allows for extremely narrow flow channels. Flows of He-II are a superposition of viscous normal flow and inviscid superflow \cite{Tilley_book}. Microscopically, the superfluid flow is irrotational. However, circulation can exist around topological defects in the superfluid, quantized vortices, which carry a single quantum of circulation $\kappa \approx 9.97\times 10^{-8}$~m$^2$s$^{-1}$ \cite{Tilley_book}. Quantized vortices facilitate friction-like coupling (mutual friction) of superfluid and normal fluid flows \cite{Barenghi1983}. On scales larger than the intervortex spacing, the flow of superfluid helium can be described using coupled equations for superflow and normal fluid flow \cite{Donnelly1991}, which in our case reduce to $\mathbf v_n \approx 0$ due to viscous clamping \cite{Rojas2015,Souris2017} for normal velocity and for superfluid velocity $\mathbf v_s$ \cite{Donnelly1991}
\begin{equation}
    \label{eq:vs-flow}
    \rho_s\frac{\mathrm D \mathbf v_s}{\mathrm D t} = -\frac{\rho_s}{\rho}\nabla p - \rho_s\alpha\kappa L \mathbf v_s + \rho_s\alpha'\omega_s\times\mathbf v_s + \mathbf f,
\end{equation}
where $\mathrm D/\mathrm D\mathrm t$ is the material derivative, $\rho$, $\rho_s$ are the total and temperature-dependent superfluid densities, respectively; $p$ is the pressure; $\alpha$ and $\alpha'$ are temperature-dependent mutual friction parameters ($\alpha = 0.034$ -- $0.181$ in the range $T = 1.3$ -- $1.85$~K \cite{Donnelly1998}, see Fig.~\ref{fig:response}a), and $L$ the vortex line density, i.e., the number of vortices per unit area; $\mathbf f$ represents the forcing. In quasi-classical cases with a large number of vortices clustered by the sign of their circulation, an approximation $|\bm{\omega}_s| = \kappa L$ can be used where $\bm{\omega}_s$ is the large-scale superfluid vorticity averaged over an area containing many vortices \cite{Skrbek2000}. Note that for a constant $L$, the $\alpha$-friction term is equivalent to large-scale friction resulting from the coupling of 2D classical flows to their 3D environment \cite{Boffetta2012}. The $\alpha'$-mutual friction resembles the Coriolis force, where the local vorticity plays the role of global rotation. Shukla \emph{et al.} have shown numerically that mutual friction affects the forward enstrophy cascade similarly to classical large-scale friction \cite{Shukla2015}.

\begin{figure}
    \centering
    \includegraphics{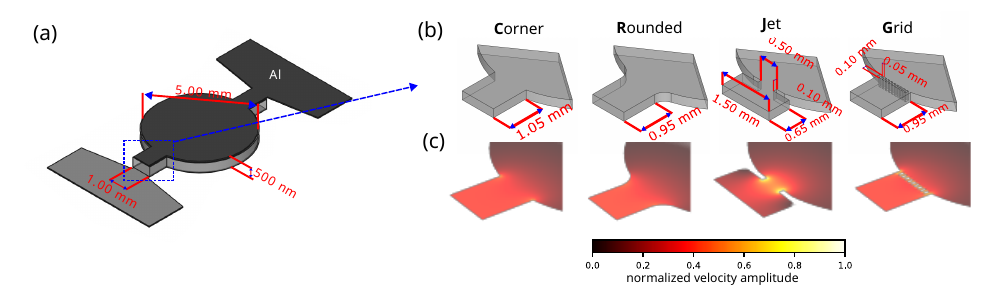}
    \caption{Nanofluidic devices. (a) Typical dimensions of the nanofluidic Helmholtz resonators. Dark-shaded areas show aluminum electrodes deposited on the top and bottom walls. The 0.5~mm quartz substrates are not shown. The entire device is submerged in superfluid $^4$He. (b) Details of the geometry of the four resonator channels used labeled C, R, J, and G (left to right) in the text. The height of all four cavities is 500 nm. The aluminum electrodes are not shown. (c) A finite element simulation shows the velocity profile of the used acoustic mode. }
    \label{fig:devices}
\end{figure}

We study two-dimensional flows using nanofluidic Helmholtz resonators, shown in Fig.~\ref{fig:devices}(a), which consist of a circular volume (``basin'') connected to the surrounding helium bath at saturated vapor pressure by two channels. Similar devices have previously been used for the study of $^3$He superfluidity \cite{Shook2020,Shook2024}, finite-size effects in $^4$He \cite{Varga2022}, quasi-two-dimensional turbulence \cite{Varga2020} and superfluid optomechanics \cite{Spence2023,Varga2021}. The superfluid helium in the channels oscillates under the restoring force provided by the compressibility of the helium and the deformation of the quartz substrate, resulting in a Helmholtz resonance \cite{Rojas2015}. The top and bottom walls of the nanofluidic volume are coated with a 50 nm aluminum layer used for capacitive actuation and sensing (see Appendix~\ref{sec:app-setup} for details on the detection circuit). A semi-open nanofluidic channel is etched onto a fused silica substrate, two of which are then bonded together to form a fully enclosed nanofluidic cavity approximately 500~nm in height (see \cite{Rojas2015,Souris2017} for further fabrication details). We studied four flow geometries with different energy injection length scales, whose velocity profile of the Helmholtz mode is shown in Fig.~\ref{fig:devices}(b,c) --- a flow around a sharp corner (device \emph{C}); a similar device with rounded-off corners (device \emph{R}); a constriction producing an oscillating jet (device \emph{J}) and a flow through a regular grid (device \emph{G}). In device \emph{R}, the behavior was similar to device \emph{C}; however, reliable velocity calibration was not possible. The data obtained from this device is available only in the attached dataset. Height $D\approx 500$~nm was used for all geometries. The characteristic length scale at which a viscous fluid follows the motion of an oscillating wall is the viscous penetration depth $\delta = \sqrt{2 \nu / \omega}$, where $\nu$ is the kinematic viscosity and $\omega$ the flow oscillation frequency. For the typical resonance frequency 2~kHz and temperature 1.65~K (see Fig.~\ref{fig:frequencies} for the temperature dependence of the resonance frequency and linewidth), and using the normal fluid kinematic viscosity from \cite{Donnelly1998}, $\delta\approx$2.7~$\mu$m, which is 5-times greater than the spacing between the upper and lower walls of the cavity. We can, therefore, consider the viscous normal fluid component of superfluid helium to be at rest. 

The typical flow velocity amplitude as a function of the drive is shown in Fig.~\ref{fig:response}(b,c). For low velocities, the scaling with the applied force remains linear, determined primarily by thermoviscous dissipation (see Appendix~\ref{sec:app-damping} and \cite{Souris2017}). Past a critical velocity, nonlinear damping sets in. The high aspect ratio of the flow indicates that the turbulence will be two-dimensional \cite{Benavides2017, Ecke2017}. However, the quantized vortex sets the smallest length scale, approximately 0.15 nm \cite{Tilley_book}. Thus, in principle, vortex loops or half-loops attached to the walls created, e.g., during reconnections, might exist in the flow. However, numerical simulations based on the vortex filament model (see Appendix~\ref{sec:app-simulations}) show that the lifetime of three-dimensional vortex configurations is significantly shorter than the flow oscillation period (several tens of microseconds, depending on the specific geometry of the loop and its orientation with respect to the flow). They are thus transient and cannot participate in long-lived turbulent structures.

\subsection{Measurement of vortex line density}
\label{sec:main-setup-vld}
Similarly to the second sound \cite{Varga2019}, the fourth sound used in this work is damped by the presence of quantized vortices. The excess nonlinear damping (shaded areas in Fig.~\ref{fig:response}(c)) can be used to calculate the average number of vortices per unit area (vortex density $L$). For completeness, we briefly reproduce the derivation of the equation of motion of the Helmholtz mode and the damping due to quantized vortices based on refs.~\cite{Varga2020,Novotny2024}.

Since only the superfluid component is moving inside the channels, the change in the density of the fluid within the basin can be expressed as 
\begin{equation}
    \delta \rho = \delta \left( \dfrac{M_B}{V_B} \right) = \dfrac{\delta M_B}{V_B} - \dfrac{M_B}{V_B^2} \delta V_B = \dfrac{1}{V_B} (-2 a \rhos y + 2 \rho A x), \label{H1}
\end{equation}
where $y$ is the displacement of the superfluid in the channels (positive for outflow), $x$ is the average displacement of the basin walls (positive for compression), the fluid mass $M_B = \rho V_B$, the volume of the basin $V_B = A D$ ($A$ is the area of the basin and $D$ its height). Changes in density and pressure are connected by isothermal compressibility $\delta \rho = \rho \chi \delta P$ and thus we can write
\begin{equation}
    \delta P = \dfrac{1}{\rho \chi V_B} (2 \rho A x - 2 a \rhos y). \label{H2}
\end{equation}
The electrostatic force acting between the charged basin plates $F = C_0 U^2/(2 D)$ ($C_0$ is the static capacitance of the chip and $U$ the total applied voltage) is equal to contributions from plate deformation and pressure
\begin{equation}
    F = k_p x + A \delta P, \label{H3}
\end{equation}
where $k_p$ is the spring constant, and we neglect the inertia of the basin walls. Combining \eqref{H1}, \eqref{H2}, and \eqref{H3}, we can write the equation of motion for the fluid displacement $y$ in the channel with length $l$
\begin{align}
    \rhos a l \ddot{y} &= \dfrac{\rhos}{\rho} a \delta P - F_f, \\
    \rhos a l \ddot{y} &= \dfrac{\rhos a k_p}{\rho^2 (\chi V_B k_p + 2A^2)} \left[ \dfrac{2A\rho}{k_p}F - 2a\rhos y \right] - F_f,\label{Hdyn}
\end{align}
where we added an additional friction force $F_f$. Here the driving force $F = C_0 U_0(t) U_B / D$, since the total voltage is a sum of the bias voltage $U_B$ and the driving alternating voltage $U_0(t)$ and we only take the component that is resonant with the Helmholtz mode. The friction force $F_f$ can be divided into two contributions $F_f = F_\mathrm{in} + F_\mathrm{ns}$. The first part is the intrinsic damping of the superfluid motion \cite{Souris2017}, which is discussed in Appendix~\ref{sec:app-damping}, and can be written in the form
\begin{equation}
    \label{H4}
    F_\mathrm{in} = a l \rhos \gamma_0 \dot{y},
\end{equation}
where $\gamma_0$ is a dimensionless intrinsic damping factor. The second contribution is the mutual friction force between the superfluid and quantized vortices, which can be expressed \cite{Donnelly1991} as
\begin{equation}
    \label{H5}
    \bv F_\mathrm{ns} = a l \alpha \rhos \hat{\bm{\omega}}_s \times [ \bm{\omega}_s \times (\bv\vn - \bv\vs) ] + a l \alpha' \rhos \bm{\omega}_s \times (\bv\vn - \bv\vs),
\end{equation}
where $\bm{\omega}_s$ is the superfluid vorticity and $\hat{\bm{\omega}}_s$ is the unit vector with the same direction. Due to geometrical constraints we have for the vortex density $L \approx |\bm{\omega}_s|/\kappa$, where $\kappa\approx 9.97 \times $10$^{-8}$ m$^2$s$^{-1}$ is the quantum of circulation. Assuming that the normal fluid component is viscously clamped to the walls, i.e. $\bv\vn = 0$, and neglecting the second non-dissipative mutual friction term, \eqref{H5} simplifies to
\begin{equation}
    F_{ns} = a l \alpha \kappa \rhos L \vs \sin^2{\theta}, \label{H8}
\end{equation}
where the angle $\theta$ is between $\bv\vs$ and $\hat{\bm{\omega}}_s$. Assuming that the vortices are preferentially oriented perpendicularly to the superfluid flow (a consequence of 2-dimensionality), we have $\sin^2{\theta} = 1$.

The dynamics of $L(t)$ likely occur on a significantly longer time scale than the flow oscillation period, as evidenced by the relatively long-lived states in both the multi-stable and intermittently switching regimes (from roughly the inverse linewidth up to several seconds; Sec.~\ref{sec:main-results-multistability}). In the steady state, we therefore consider $L$ to be a constant. This constant will, however, depend on the RMS flow velocity (and thus drive amplitude and frequency). Therefore, a Lorentzian peak shape of the resonant response should not be expected. A similar constancy of $L$ generated by the second sound in a 3D geometry was reported in turbulence generated by the second sound \cite{Midlik2021}.

Identifying $\dot{y}$ with $\vs$ and substituting friction \eqref{H4},\eqref{H8} into the equation of motion \eqref{Hdyn} results in
\begin{equation}
    \label{H6}
    \ddot{y} + (\alpha \kappa L + \gamma_0) \dot{y} +\dfrac{2 \rhos a k_p}{l \rho^2 (\chi V_B k_p + 2 A^2)} y = \dfrac{2AC_0U_B}{\rho l (\chi V_B k_p + 2A^2) h} U_0(t),
\end{equation}
which has the form of the equation of motion of a linear harmonic oscillator
\begin{equation}
    \ddot{y} + \Tilde{\gamma} \dot{y} + \omega^2_0 y = f_0 e^{i\omega t},
\end{equation}
with the total damping factor $\Tilde{\gamma}$, the resonance frequency $\omega_0$ and a driving force $f_0(t)$
\begin{align}
    \Tilde{\gamma} &= \alpha \kappa L + \gamma_0 = \gamma (L) + \gamma_0, \\
    \omega_0^2 &= \dfrac{2 \rhos a k_p}{l \rho^2 (\chi V_B k_p + 2A^2)},\\ \label{res freq}
    f_0(t) &= \dfrac{2 A C_0 U_B}{\rho l (\chi V_B k_p + 2A^2)h} U_0(t) = f_0 e^{i \omega t}.
\end{align}

On resonance, $\omega=\omega_0$ the complex response is given in the vortex-free state ($L=0$) by $y_0 = f_0/(i \omega_0 \gamma_0)$ and for $L>0$ $y = f_0/(i \omega_0 (\gamma(L) + \gamma_0))$, which yields
\begin{equation}
    \dfrac{y_0}{y} - 1 = \dfrac{\gamma(L)}{\gamma_0}.
\end{equation}

Using Eq.~\ref{H8} (with $\theta = \pi /2$) and the relation between the damping factor $\gamma_0 $ and the width $\Delta_0$ of the un-damped resonance curve, $\gamma_0 = 2 \pi \Delta_{0}$, results in
\begin{equation}
    \label{eq:VLD}
    L = \dfrac{2 \pi \Delta_0}{\alpha \kappa} \left( \dfrac{y_0}{y} - 1\right).
\end{equation}

In the present work, $y_0$ is extrapolated from the linear regime of the force-velocity curves shown in Fig.~\ref{fig:response}(c).

\section{Results}
\label{sec:main-results}

To observe the development of turbulence, the Helmholtz resonance was excited on its resonance frequency with slowly swept drive amplitude in increasing and decreasing directions. The typical response showing the excess dissipation past a critical velocity is shown in Fig.~\ref{fig:response}(c). Since the flow can transition stochastically between several states, the drive sweep is repeated in several cycles to sample the possible states fully.

\subsection{First critical velocity}
\label{sec:main-results-vc1}

\begin{figure}
    \centering
    \includegraphics{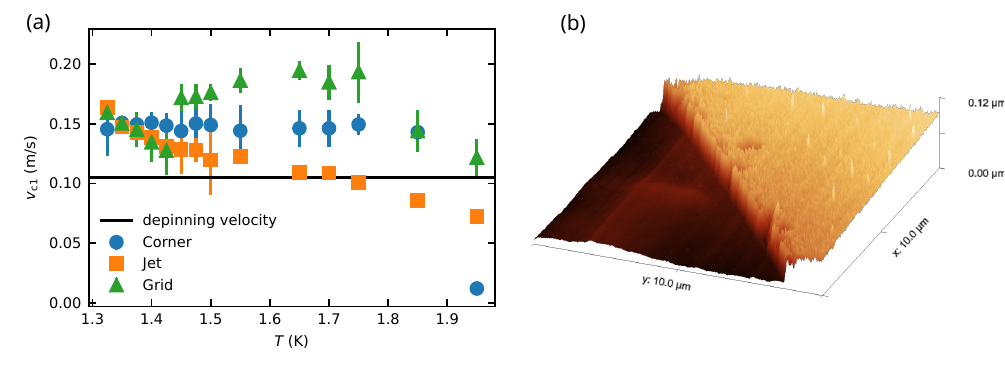}
    \caption{First critical velocity. (a) The first critical velocity for the three device geometries. Within a factor of two, the velocities agree with the depinning velocity estimate \eqref{eq:depin}. The critical velocity is determined as the velocity, where vortex line density increases above $5 \times 10^8$~m$^{-2}$ and the ends of the errorbars as velocities where $(5\pm 2.5) \times 10^8$~m$^{-2}$ is exceeded. (b) Atomic force microscopy scan of the topography of the edge of the aluminum surface (higher, rougher area) in the basin area.}
    \label{fig:vc1}
\end{figure}

As the flow velocity increases, the nonlinear damping first appears at approximately 0.1 -- 0.2 m/s (see appendix~\ref{sec:app-calibration} for calibration of the measured electrical quantities to forces and velocities). For concreteness, we define the critical velocity as the velocity where the vortex line density calculated using \eqref{eq:VLD} first exceeds a threshold of $L_c = 5 \times 10^8$~m$^{-2}$ with results shown in Fig.~\ref{fig:vc1}(a). The velocities are only weakly dependent on device geometry and temperature, and the transition is generally reversible (with the exception of device \emph{G} at the lowest attained temperatures, see Fig.~\ref{fig:bifurcation}).

A possible geometry and temperature-independent mechanism for the appearance of critical velocity is vortex pinning. Due to the small size of the vortex core, $a_0\approx 0.15$~nm, vortices will be pinned on surface roughness and will be unable to slide unless subject to superflow faster than the depinning velocity, which was estimated by Schwarz \cite{Schwarz1985} for a semi-spherical protuberance of radius $b$ on a planar surface in a planar channel of height $D$ as
\begin{equation}
    \label{eq:depin}
    v_\mathrm{pin} = \frac{\kappa}{2\pi D}\ln\left(\frac{b}{a_0}\right).
\end{equation}

Because of the stringent requirements of the surface cleanliness for the bonding of the resonators, any manipulation of the unbonded chips outside of the cleanroom environment is essentially destructive, therefore we were unable to measure the topography on the actual devices used in the study. However, a scan of the aluminum layer surface topography using atomic force microscopy is shown in Fig.~\ref{fig:vc1}(b) on devices created using a nearly identical fabrication process. The scan shows 10 \textmu m square on the edge of the aluminum. The step height is 48.64 $\pm$ 0.04 nm, and the RMS surface roughness (that is, the root-mean-square distance to the mean plane) of the aluminum layer is approximately 4.1 nm.

Replacing the protuberance size $b$ in \eqref{eq:depin} with the measured surface roughness results for $D=500$~nm in $v_\mathrm{pin} \approx $ 11 cm/s, which is in good agreement with initial reversible transitions to nonlinear dissipation seen in Fig.~\ref{fig:vc1}(a). The origin of the slight temperature dependence of this critical velocity is unknown. However, \eqref{eq:depin} was obtained for steady flow for a single pinning site; in the present case, the temperature dependence is probably the result of the interaction of the temperature-dependent flow oscillation frequency and the vortex relaxation time given by the mutual friction \cite{Barenghi1985}.

From the point of view of classical fluid turbulence, the typical critical velocities $v_c = $ 0.1 to 0.2 m/s correspond to Reynolds numbers defined with channel height $D$, $\mathrm{Re} = UD\rho/2\eta =$ 2 to 5 for viscosity $\eta\approx 10^{-6}$ Pa$\cdot$s \cite{Donnelly1998}. This is below the onset of spatiotemporally intermittent turbulence in planar shear flows for which the critical $\mathrm{Re}_c\approx 330$ \cite{Klotz2022}. However, for $\mathrm{Re}$ defined using in-plane geometry (i.e., lateral channel width) $\mathrm{Re}_c \approx 10^3$ to $10^4$ indicating turbulent flow with a developed inverse cascade \cite{Xia2014}.

\subsection{Hysteresis and multistability in turbulent transition}
\label{sec:main-results-multistability}

\begin{figure}
    \centering
    \includegraphics{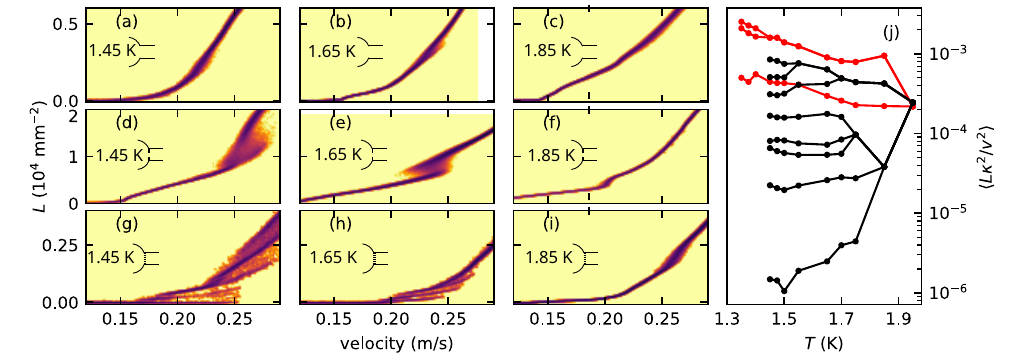}
    \caption{Vortex density bifurcation diagrams. (a--i) Two-dimensional histograms of vortex density and velocity (darker color means higher probability; logarithmic color scale) for device type \emph{C} (panels a--c), \emph{J} (d--f) and \emph{G} (g--i) for three temperatures, 1.45 K (a, d, g), 1.65 K (b, e, h), and 1.85 K (c, f, i). (j) The temperature-density bifurcation diagram shows the average value of $L\kappa^2/v^2$ calculated for each individual branch of $L(v)$ (averaged over drives). Red, device \emph{J}, black, device \emph{G}. For $T < 1.45$~K, the data for device \emph{G} continue to branch in a fractal manner.}
    \label{fig:bifurcation}
\end{figure}

The vortex density calculated from the excess damping is shown in Fig.~\ref{fig:bifurcation}(a--i) as two-dimensional histograms for three geometries for a high, intermediate, and low temperature (1.85~K, 1.65~K and 1.45~K, respectively), which determines the large-scale friction in \eqref{eq:vs-flow}. Data from all devices and 14 temperatures in total are available in the attached dataset.

At high temperatures above a critical velocity, the observed vortex density gradually and reversibly increases above zero for all geometries. For lower temperatures, the behavior strongly depends on geometry. For device \emph{C} (panels (a--c)), weak hysteresis is observed after the first critical velocity. For device \emph{J} (panels (d--f)), the flow transitions to a regime with increased $L$ through a discontinuity with stochastic transitions between two discrete states. For device \emph{G} (panels (g--i)), we observe several bifurcations below approximately 1.65 K, resulting in multiple metastable states and hysteresis in the transition region. Additionally, at least two intermittent transitions are observable in panel (g). In Fig.~\ref{fig:bifurcation}(j), we show the bifurcation diagram of $L$ as a function of temperature. Each branch separated by discontinuous transitions from the others is characterized by a dimensionless quantity $\langle L\kappa^2/ v^2\rangle$ (averaging over drive amplitudes within a given branch). For device \emph{J}, two branches are observed from 1.85~K to 1.4~K, where the upper branch divides into two. For device \emph{G}, the branches proliferate in a fractal fashion below 1.45~K. However, the growth in complexity results in a less reliable classification of the data.

\begin{figure}
    \centering
    \includegraphics{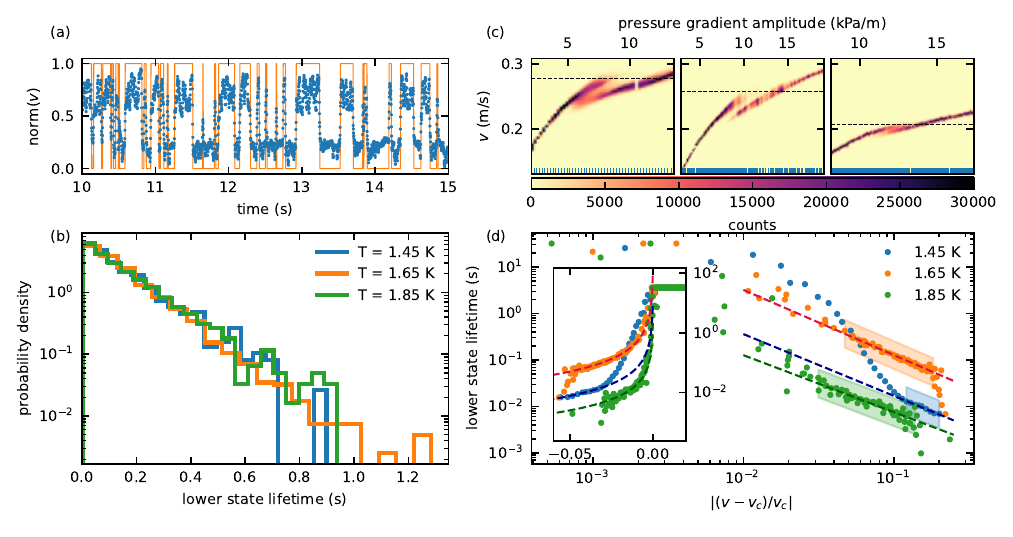}
    \caption{Critical behavior of turbulent transition in device \emph{J}: (a) Time dependence of the observed velocity amplitude in the intermittently switching region at 1.65 K at drive pressure gradient approximately 9.5 kPa/m. Maximum and minimum velocity were normalized to 0 and 1, respectively; the orange line shows the binarized signal used for the detection of transitions. (b) Histrogram of the turbulent state lifetimes in the transitional region at the three indicated temperatures for datasets with a mean lifetime in the range 0.1 to 0.2 s. A straight line in semilogarithmic axes indicates exponentially decaying dependence. (c) Histograms of observed velocity amplitude for a fixed drive at 1.45 K, 1.65 K, and 1.85 K (left to right). The rug plot indicates the used drives. (d) The mean lifetime of the lower (more dissipative) state as a function of dimensionless reduced velocity in that state for the three temperatures (the critical velocity is indicated in the histograms by horizontal dashed lines). The power law (dashed lines) was fit to the range indicated by the shaded areas. (inset) Same data against dimensional shifted velocity $v - v_c$ (in m/s).}
    \label{fig:lifetimes}
\end{figure}

Two possibilities are currently available for the nature of discrete turbulent states: critical transition between forward and inverse cascades (e.g., \cite{Alexakis2023}) or organization of the vorticity into condensates. Due to the existence of more than two discrete states, the latter option appears more likely. Thus, assuming that the vortex line density can cluster in several distinct distributions (``modes'', e.g., large-scale vortices or zonal jets \cite{Xu2024}) we seek a low-dimensional model of the vortex density $L$ (similarly to ref.~\cite{Varga2020}) to describe the appearance of the discontinuous transitions in devices \emph{J} and \emph{G}.  The behavior can be qualitatively captured with a dynamical system (for dimensionless quantities)
\begin{equation}
    \label{eq:L-dynamics-multi}
    \frac{\mathrm d L_j}{\mathrm d t} = -d_l L_j - \left(\frac{L_j}{1 + L_j^2}\right)^2 + g_j,
\end{equation}
where $L_j$ is the vortex density in mode $j$, $j=1...N$, $N$ is the number of participating modes,  $d_l$ represents the strength of the large-scale damping due to mutual friction (first term) relative to the vortex pair annihilation (second term), and $g_j$ represents the forcing, which, in general, will be a nonlinear function of the flow velocity and geometry. For the total vortex density, we take $L^2 = \sum_j L^2_j$. The essential feature of the model is that, for large populations $L > 1$, the annihilation of vortices is suppressed. A typical curve $\mathrm dL/\mathrm dt$ for $N=1$ is shown in Fig.~\ref{fig:lifetimes}(a), and a bifurcation diagram with drive $g$ as the control parameter is shown in Fig.~\ref{fig:lifetimes}(b). For low damping $d_l$, the time derivative exhibits a local minimum, resulting in three fixed points, two stable and one unstable, for a certain range of $g$, which qualitatively describes the bifurcation seen in device \emph{J}. With increasing temperature, the mutual friction increases, increasing large-scale friction on the vortex system. Indeed, the bistability is lost when the parameter $d_l$ is increased, similarly to the experiment (cf. Fig.~\ref{fig:lifetimes}(b) and Fig.~\ref{fig:bifurcation} middle row).

The energy injection scale $l_f$ for the device \emph{J} is close to the width of the constriction, about half of the maximum size of the system $l_0$. Therefore, only a small number of wave vectors can exist in the inverse cascade between $l_f$ and $l_0$. In device \emph{G}, we increase the ratio $l_0/l_f$, thus extending the number of modes participating in the inverse cascade. Therefore, we set $N>1$ and assume that the forcing is localized at $j=1$, $g_1 = g$ and that the higher modes are forced in a cascading manner $g_j = cL_{j-1}$,  $j>1$, where $c$ characterizes the mode coupling ($d_l$ and $c$ are kept constant for simplicity). For $d_l=0.05$, $c=0.04$, the total density $L$ in Fig.~\ref{fig:lifetimes}(c) shows multistability due to overlapping pairs of saddle-node bifurcations resembling the experimental data of device \emph{G} (full lines show stable stationary solutions; dashed lines the unstable fixed points). Although the coupling between vorticity modes is simplified \cite{Kraichnan1980}, multistable behavior can occur for more complicated multiplicative coupling \cite{Huang2016}. Extending the interpretation of the single-mode dynamics of device \emph{J}, the resulting physical picture is that, as a particular mode is sufficiently excited, the vortices cluster in a certain pattern to reduce pair annihilation. The energy in the mode thus increases, which allows nonlinear interactions to force progressively larger scales until the size of the system is reached. The suppression of the annihilation is responsible for the appearance of a pair of bifurcations and the bistable regime. The exact form of the annihilation term does not affect the qualitative behavior of the solutions as long as a local minimum appears. The rational annihilation term in \eqref{eq:L-dynamics-multi} was chosen to maintain the expected $L^2$ behavior for low $L$ corresponding to the annihilation of randomly distributed vortices \cite{Varga2018}. 

Note that the bistability observed in device \emph{J} is different than observed previously in 1~\textmu m confinement \cite{Varga2020}, where it was long-lived on the time scale of the experiment (i.e., minutes). Here we observe intermittent switching or pronounced multistability. However, present results continue the trend observed in \cite{Varga2020} of weakened bistability for \emph{C}-type device as confinement $D$ decreases. Both types of multi-stability should be considered as special cases of more general dynamical behavior, a complete understanding of which is not yet available.

\begin{figure}
    \centering
    \includegraphics{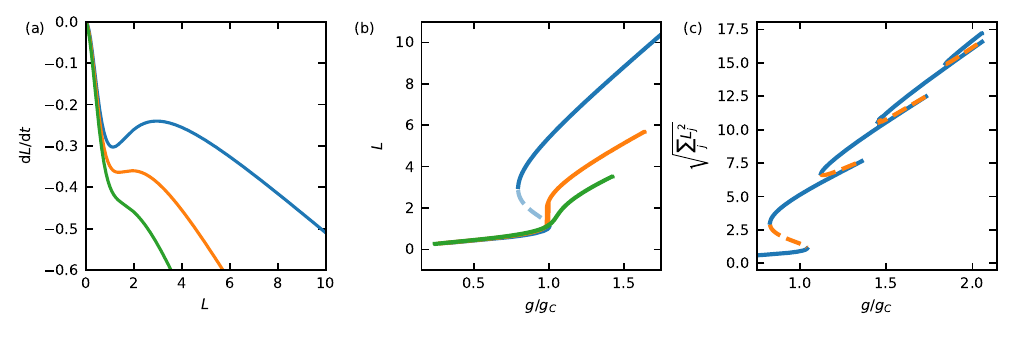}%
    \caption{Nonlinear dynamical model of quasi-2D turbulence: (a) The single-mode model in Eq. \eqref{eq:L-dynamics-multi} for $N=1$, $d_l=$ 0.05, 0.1, and 0.15 (top to bottom curve) and $g=0$, $g>0$ shifts the curve along the y-axis. (b) Stable (full lines) and unstable (dashed lines) stationary solutions of \eqref{eq:L-dynamics-multi} as a function of $g$ (normalized by the position of the local minimum or the inflection point $g_c$ of \eqref{eq:L-dynamics-multi}) for the same values of $d_l$ as in panel (a). (c) Multistability with several overlapping pairs of saddle-node bifurcations develops by coupling multiple modes of vortex density in a cascading manner as in \eqref{eq:L-dynamics-multi}, $N=4$. The orange dashed lines show unstable stationary solutions. }
    \label{fig:enter-label}
\end{figure}

\subsection{Stochastic transitions and critical behavior.}
\label{sec:main-results-transitions}

In the region of intermittent transitions in device \emph{J} it is apparent that the the probability of finding the turbulence in the more dissipative state gradually increases as the velocity is increased. To capture the statistics of the intermittent transitions for device \emph{J} we recorded the time series of the response amplitude for 1.45 K, 1.65 K, and 1.85 K at a set of fixed drive amplitudes. An example is shown in Fig.~\ref{fig:lifetimes}(a), which shows clear switching between distinct states; records were 32~s long, i.e., capturing over 60000 flow oscillation periods, which are about 0.5~ms long (i.e., 2 kHz resonance frequency, Fig.~\ref{fig:response}(b)). The resulting two-dimensional histogram in Fig.~\ref{fig:lifetimes}(c) shows a bimodal shape equivalent to the ramped drive data in Fig.~\ref{fig:bifurcation}. The observed time series was threshold-binarized to detect transitions which were then used to calculate the lifetimes of the more dissipative (lower velocity) state as the average time between the system jumping into and out of the state. The probability density function of the state lifetime is exponential (see Fig.~\ref{fig:lifetimes}(b)), suggesting that the transitions are memoryless. The mean lifetime $\tau$ as a function of the velocity, Fig.~\ref{fig:lifetimes}(d), shows a diverging behavior consistent with a power law of the form $\tau \propto (v_c - v)^{-\eta}$ with $\eta = 1.35 \pm 0.07$. The exponent was calculated as the mean of the three temperatures shown in Fig.~\ref{fig:lifetimes}(b) weighted by fit error determined by bootstrap \cite{Efron1981}. We note, however, that the scaling is observed only for approximately half a decade for the two higher temperatures due to a limited number of transitions very close to the critical point. For 1.45~K, we observe only a possible indication of the scaling restricted to $(v_c - v)/v_c > 0.1$; close to the critical point the transitions become less distinguishable.

\section{Discussion and conclusions}
\label{sec:main-discussion}

The critical exponent of the stochastic fluctuating transition is close to the longitudinal correlation critical exponent of 2+1D directed percolation $\xi_{||} \approx 1.295$ which describes transition to turbulence in shear flows \cite{Avila2011,Manneville2016,Hof2023} although the present layer-height critical $\mathrm{Re}\approx 5$ is significantly smaller than that of classical shear flows $\mathrm{Re}_c\approx 330$ \cite{Klotz2022}. The mechanism for transition to turbulence is, therefore, most likely different between the oscillating superflow and classical shear flows.

The essential feature of our dynamical model \eqref{eq:L-dynamics-multi} describing the formation of multistability is the decrease in the rate of vortex annihilation at high densities, which is consistent with clustering of vortices of equal signs which occurs during vorticity condensation \cite{Gauthier2019,Sachkou2019}. The intermittent transitions seen in Fig.~\ref{fig:lifetimes}(a) resemble the stochastic transitions between large-scale vortices and zonal jets \cite{Xu2024,Bouchet2009} or between multiple zonal jet configurations \cite{Simonnet2021} observed in several numerical works (including three-state multistability observed in a simulation of anisotropic 2D turbulence by Xu \emph{et al.}~\cite{Xu2024}).  However, Xu \emph{et al.} observed that the lifetimes of both types of condensates increase exponentially with increasing Reynolds number, whereas we observe a diverging increase in the lifetime of one state at the expense of the other with a critical exponent. Interestingly, the reversible-irreversible transition in periodically driven many-body systems \cite{Reichhardt2023} shows a similar critical exponent, about 1.33 \cite{Corte2008}, in the time to reach a steady state close to the transition. This suggests that the switch in the preferred turbulent state as flow parameters are varied falls within the universality class of this nonequilibrium phase transition, although measurement of the state lifetimes over a broader range of reduced velocities is needed for a more reliable estimate of the critical exponent. Superficially the 

Relaxing the assumption of strict two-dimensionality, the condensate was found to form discontinuously at a critical forcing scale to layer thickness ratio \cite{Kan2019a,Kan2019,Wit2022,Alexakis2023}. The geometry of our flow is fixed; however, the typical length scale over which vortices move will rapidly increase upon exceeding the depinning velocity \eqref{eq:depin}, which will result in the increase of the energy injection scale. Indeed, the bistability we observe is similar to scenario 3 of ref.~\cite{Kan2019}, where the formation of 2D inverse cascade and cessation of 3D direct cascade occur at distinct critical points with a bidirectional cascade in the intermediate range, although no power-law divergence was observed there. We are, therefore, unable to definitively identify the nature of the multiple turbulent states and the transitions between them since none of the currently available numerical results fully account for the observed multistability and the statistics of transitions.

Finally, we note in passing that the aspect ratios of the studied flows are comparable to geophysical flows, where self-organizing behavior is common \cite{Alexakis2024,Marcus2004,Yadav2020} and thus, the presented results might help shed light on transitions and formation of large-scale structures in planetary flows. Indeed, baroclinic turbulence was shown to be well described by a gas of point vortices \cite{Gallet2020,Meunier2025} and multistability is present in the climate model \cite{Margazoglou2021}, e.g. in the simulations of large-scale flows of the Atlantic ocean \cite{Ditlevsen2023,vanWesten2024} in context of the collapse of the Atlantic meridional overturning circulation, which would have significant impact on the global climate.

\section*{Data Availability}
All data and software needed to generate the figures in the paper are available in Zenodo repository at \url{doi.org/10.5281/zenodo.14824085}, which also contains tools to view all underlying data of the histograms shown in Figs.~\ref{fig:bifurcation} and \ref{fig:lifetimes}.

\begin{acknowledgments}
    We would like to thank K. Olejník for providing access to the clean rooms at FZU and K. Uhlířová for AFM characterization. The authors also thank L. Skrbek, P. \v{S}van\v{c}ara and R. Dwivedi for fruitful discussions. The work was supported by Charles University under PRIMUS/23/SCI/017. CzechNanoLab projects LM2023051 and LNSM-LNSpin funded by MEYS CR are gratefully acknowledged for the financial support of the sample fabrication at CEITEC Nano Research Infrastructure and LNSM at FZU AV\v{C}R. AFM characterization was performed in MGML laboratories supported by the program of Czech Research Infrastructures (project no. LM2023065).
\end{acknowledgments}

\appendix

\section{Drive and detection circuitry}
\label{sec:app-setup}

\begin{figure}
    \centering
    \includegraphics[width=\textwidth]{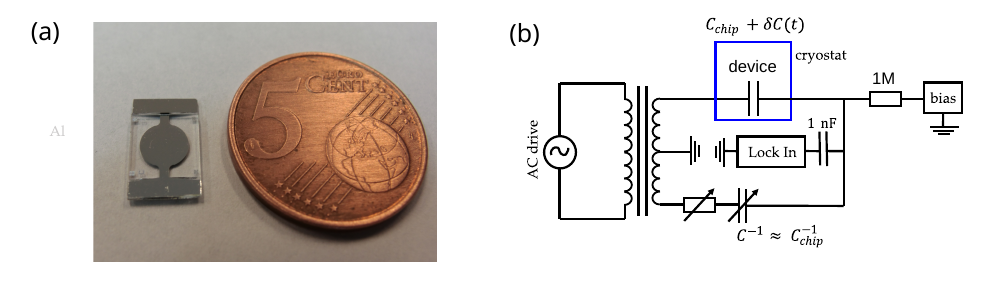}
    \caption{Nanofluidic resonator and detection circuit. (a) A photograph of an assembled Helmholtz resonator used for measurement next to a 5 cent Euro coin for scale. (b) The capacitance bridge circuit used for the drive and detection of the flow. The center-tapped transformer excites both sides of the bridge circuit with an alternating voltage with $\pi$ phase shift between the bridge arms. When the bridge is tuned, only the current resulting from the oscillation of the device capacitance flows through the lock-in amplifier. In order to avoid measuring at the second harmonic (since the electrostatic force is proportional to voltage squared), the device is biased by (typically) a 10 V DC voltage. }
    \label{fig:circuit and chip}
\end{figure}

The flow in the resonator is driven by two aluminum electrodes deposited opposite each other inside the cavity. An applied alternating voltage causes deformation of the substrate, leading to pressure oscillation inside the cavity and to a pressure gradient along the channels. The resulting motion of the superfluid causes a fluctuation in the capacitance of the device $\delta C(t)$. This fluctuating capacitance is detected using a bridge circuit shown in Fig. \ref{fig:circuit and chip}(b). The capacitance of the lower branch $C$ is adjusted to the static capacitance of the device $C_0$ (typically about 400~pF). Sufficiently far from the Helmholtz resonance, zero current is measured by the lock-in amplifier when the bridge is tuned. As the capacitance fluctuations increase close to the resonance, the charge on the device starts to oscillate due to the bias, which is detected by the lock-in amplifier as an oscillating current.

\begin{figure}
    \centering
    \includegraphics[width = \textwidth]{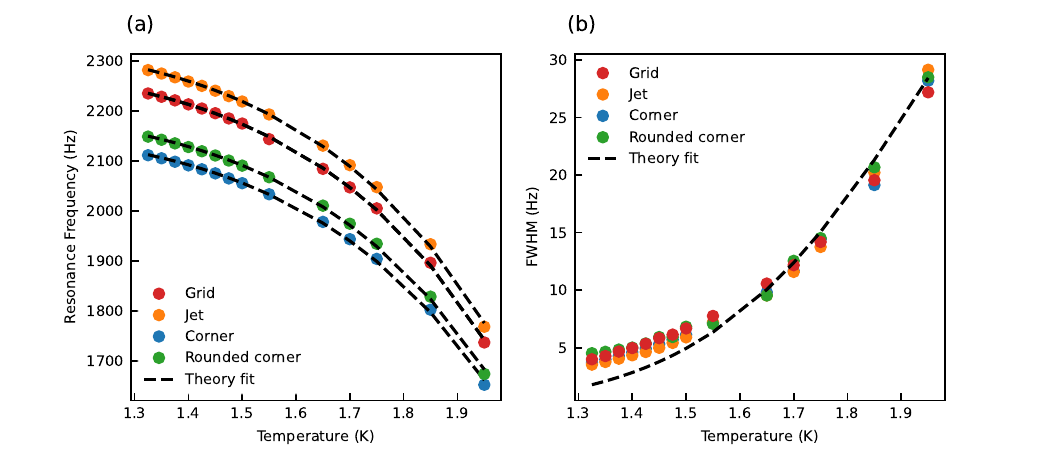}
    \caption{(a) Resonance frequencies for temperatures between 1.325~K and 1.950~K for the four used chip geometries. The dashed line is fit to \eqref{eq:corr_res_freq} considering the temperature correction; for the fit parameters, see Tab. \ref{tab:parameters}. (b) The corresponding widths (FWHM) of the resonances. The width is equal to the linear part of the damping of the resonance motion. The black dashed line is a fit of the sum of \eqref{delta n} and \eqref{delta th}.}
    \label{fig:frequencies}
\end{figure}

The frequency $f_0$ of the Helmholtz resonance is given by \cite{Souris2017}
\begin{equation}
    \label{eq:corr_res_freq}
    f_0 = \dfrac{1}{2 \pi}\sqrt{c \dfrac{ \rhos a k_p}{l \rho^2 A^{2} (1 + \Sigma)}},
\end{equation}
where $\Sigma = (\chi D k_p)/(2 A)$, $A$ is the area of the basin, $a$ the cross-sectional area of the channel and $l$ its length, and $\chi$, $\rho$ and $\rho_s$ are isothermal compressibility, and total and superfluid densities. Compresibility values $\chi$ are taken from \cite{Brooks1977}, the total density $\rho$ and the superfluid density $\rhos$ from \cite{Donnelly1998}. The parameter $c$ is one of the geometric correction parameters introduced by Souris \emph{et al.} \cite{Souris2017} that corrects for the finite flow in the basin. The measured resonance frequencies for the four devices, along with a fit to \eqref{eq:corr_res_freq}, are shown in Fig.~\ref{fig:frequencies}. Furthermore, we find an improvement of the fit if we allow for a small systematic error in the temperature calibration as $T_r = bT$, where $T$ is the measured temperature, $T_r$ is the actual temperature, and $b$ is the correction parameter. The correction parameters are shown in Tab.~\ref{tab:parameters}.

\begin{table}
    \centering
    \begin{tabular}{c|c|c|c|c|c}
         Chip Type & $b$ & $c$ & $\lambda$ & $k_p (10^7 \textrm{N/m})$ & $C_0 (\textrm{pF})$\\
         \hline
         \hline
         Corner & 0.97 & 1.77 & 1.94 $\pm$ 0.08 & 1.24 & 338.6\\
         \hline
         Rounded corner & 0.97 & 2.54 & 2.05 $\pm$ 0.07 & 1.19 & 312.9 \\
         \hline
         Jet & 0.97 & 1.82 & 2.07 $\pm$ 0.10  & 1.45 & 335.7 \\
         \hline
         Grid & 0.97 & 2.78 & 1.77 $\pm$ 0.09 & 1.62 & 466.4\\        
    \end{tabular}
    \caption{Device parameters. Values of correction parameters obtained by fitting Eq. \eqref{eq:corr_res_freq} in case of $b$ and $c$, and the sum of Eq. \eqref{delta n} and \eqref{delta th} in the case of $\lambda$. The error of $b$ - correction factor is $\pm$ 10$^{-3}$ and of $c$ - correction factor $\pm$ 2$\times$10$^{-3}$. The static capacitances $C_0$ and spring constants $k_p$ for all four used geometries were calculated from parameters of a linear fit to equation \eqref{C3}. The error of the spring constant is $\pm$ 0.01$\times$10$^{7} \textrm{N/m}$ and of the static capacitance $\pm$ 10$^{-2}$pF.} 
    \label{tab:parameters}
\end{table}

\section{Calibration of force and velocity}
\label{sec:app-calibration}

The voltage between the device electrodes and the resulting current differ from the voltage applied to the primary winding of the center-tapped transformer and the current measured by the lock-in amplifier due to circuit losses. These losses need to be taken into account to obtain accurate velocities and driving pressure gradients. The frequency and applied voltage dependence of the circuit losses were characterized by directly measuring the oscillating voltage across the device with a separate lock-in amplifier. The current losses were characterized by injecting a known oscillating current immediately after the device on the bias side (see Fig.~\ref{fig:circuit and chip}(b)). The resulting transfer functions were used to correct the driving force (proportional to the voltage across the device) and the flow velocity (proportional to the detected current).

The calculation of the mechanical quantities also requires the effective spring constant of the deformation of the substrate in the basin region. Let $C_0$ be the capacitance of the device corresponding to the zero DC bias voltage $U_B$. We characterize the effective spring constant $k_p$ by measuring the dependence of the device capacitance on the applied bias voltage $U_B$ similarly to ref.~\cite{Souris2017}. The force acting on the basin plates is
\begin{equation}
    \label{C1}
    F = \dfrac{CU^2}{2D} = k_p x,
\end{equation}
where $D$ is the distance between the electrodes, $x$ is the mean vertical deformation of basin plates and for the capacitance $C$, we have
\begin{equation}
    \label{C2}
    C = \dfrac{A \varepsilon_0}{D - 2x} \approx \dfrac{A \varepsilon_0}{D} + \dfrac{2 \varepsilon_0 A}{D^2}x  = C_0 + \dfrac{2C_0}{D}x,
\end{equation}
where $A$ is the basin area, $\varepsilon$ is the vacuum permittivity, and we assumed  $x \ll D$, taking into account only the first two terms of Taylor expansion. Furthermore, $x$ can be expressed from Eq. \eqref{C1} and substituting into Eq. \eqref{C2} we get
\begin{equation}
    \label{C3}
    C = C_0 + \sigma U^2,
\end{equation}
where $\sigma = C_0^2/(D^2 k_p)$. 

\begin{figure}
    \centering
    \includegraphics[width = \textwidth]{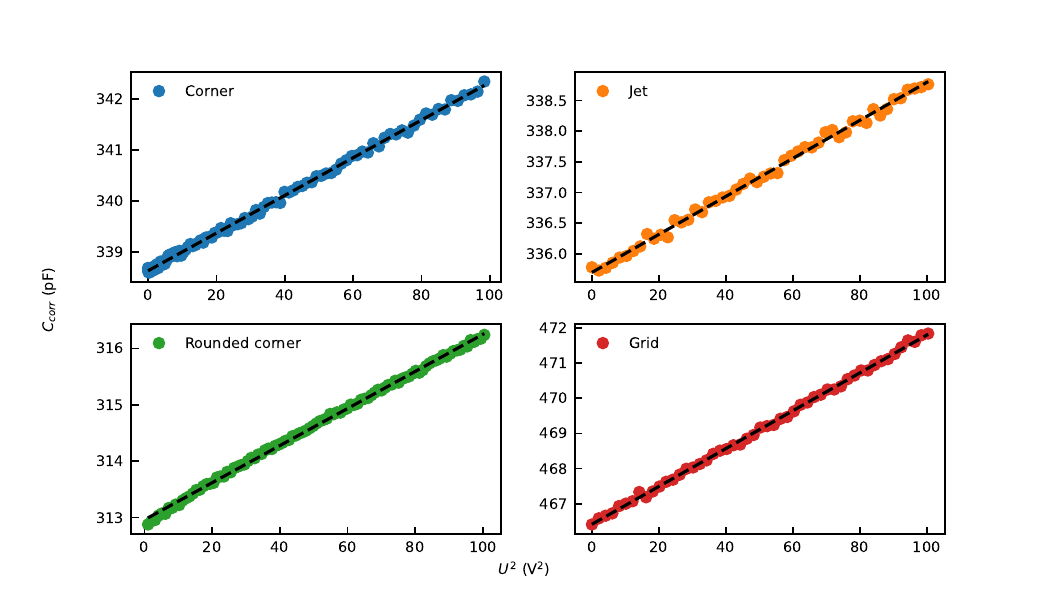}
    \caption{The corrected capacitance of all four Helmholtz resonators against the applied voltage. Only the DC bias voltage amplitude was changed, and the amplitude of the AC drive was kept constant at 0.02~Vrms. The black dashed line is a fit of the Eq. \eqref{C3}. From the fit parameters we calculate $C_0$ and $k_p$ and these values can be found in the Tab. \ref{tab:parameters}.}
    \label{fig:kp_calib}
\end{figure}

Measured capacitance includes the capacitance of the electrodes in the flow channels $C_\mathtt{channel}$ and a 100~nF isolation capacitor connected in series. The corrected basin capacitance is
\begin{equation}
    \label{S9}
    C_{corr} = 1/(1/C_{meas} - 1/100\textrm{nF}) - 2 C_\mathtt{channel}.
\end{equation}
The corrected $C_{corr}$ as a function of the applied voltage squared $U^2$ is shown in Fig.~\ref{fig:kp_calib}. The data fit with the Eq. \eqref{C3} to obtain $C_0$ and $k_p$, shown in Tab.~\ref{tab:parameters}

% \begin{table}
%     \centering
%     \begin{tabular}{c|c|c}
%          Chip Type & $k_p (10^7 \textrm{N/m})$ & $C_0 (\textrm{pF})$ \\
%          \hline
%          \hline
%          Corner & 1.24 & 338.6 \\
%          \hline
%          Rounded corner & 1.19 & 312.9 \\
%          \hline
%          Jet & 1.45 & 335.7 \\
%          \hline
%          Grid & 1.62 & 466.4 \\        
%     \end{tabular}
%     \caption{Obtained static capacitances $C_0$ and spring constants $k_p$ for all four used geometries. Values are calculated from parameters of a linear fit, representing the equation \eqref{C3}. The error of the spring constant is $\pm$ 0.01$\times$10$^{7} \textrm{N/m}$ and of the static capacitance $\pm$ 10$^{-2}$pF.}
%     \label{tab:C0 and kp}
% \end{table}

The relevant mechanical parameters are the pressure gradient $\delta P/l$ along the flow channel resulting from the electrostatic driving force and the amplitude of the flow velocity. The equation for $\delta P/l$ can be obtained from the driving term in \eqref{H6}
\begin{equation}
    \label{gradP}
    \dfrac{\delta P}{l} = \dfrac{2AC_0U_B}{l(\chi V_B k_p + 2A^2)h} U_0,
\end{equation}
where $U_0(t)$ was replaced by its amplitude $U_0$. Based on ref.~\cite{Varga2020}, velocity is calculated as
\begin{equation}
    \label{velo}
    \vs = \dfrac{I}{g U_B C_0},
\end{equation}
where $I$ is the detected current (corrected for losses) and $g$ is defined as 
\begin{equation}
    \label{eq:velocity-g}
    g = \dfrac{2 a \rhos}{V_B \rho} \left(1 - \dfrac{2(\varepsilon - 1)}{\varepsilon} \right) \dfrac{1}{1 + 2 \Sigma},
\end{equation}
where $\varepsilon$ is the dielectric constant taken from \cite{Brooks1977}, and  $\Sigma = \chi D k_p/(2 A)$.

\section{Fourth sound damping}
\label{sec:app-damping}

Following the analysis of refs.~\cite{Backhaus1997,Souris2017}, linear damping of the Helmholtz resonance \eqref{H4} in the temperature range 1.3 K to 2 K has two dominant contributions: viscous normal fluid motion and thermoviscous dissipation. For the inverse of resonance width $\Delta_n$ caused by the viscous normal flow we can write \cite{Souris2017}
\begin{equation}
    \label{delta n}
    \Delta_n^{-1} = 2\pi\dfrac{8 \eta}{(D/2)^2}\dfrac{\rhos}{\rhon^2}\dfrac{1}{\omega_0^2},
\end{equation}
where $\eta$ is the dynamic viscosity.

The second contribution is due to the heat transfer (resulting in energy loss of the oscillating flow) from the basin to the surroundings (for further details see ref.~\cite{Backhaus1997}). The temperature gradient between the basin and the bulk $^4$He is a consequence of a mechanocalorical effect, that is, the change in the superfluid density inside the basin results in a change in temperature. This effect can be characterized by the parameter
\begin{equation}
    \Gamma^2 = \dfrac{\rho^2 s^2 T A^2 (1 + \Sigma)}{C_\mathrm{th} k_p}, \label{delta_n}
\end{equation}
where $C_\mathrm{th} = C_s V_B$ is the total heat capacity of $^4$He inside the basin ($C_s$ is the specific heat capacity of $^4$He) and $s$ is the specific entropy. For the case $\Gamma^2 \ll 1$ (in our case $\Gamma \approx 0.22$ at 1.65 K), the influence of the additional pressure gradient caused by the presence of the temperature gradient (i.e., the fountain pressure) is negligible and for damping due to the heat flux out of the basin, it can be shown that \cite{Souris2017}
\begin{equation}
    \label{delta th}
    \Delta_{th}^{-1} = \dfrac{1 + \Phi^2}{\Phi \Gamma^2}\left( 1 + \dfrac{\Gamma^2}{2(1 + \Phi^2)} \right)\dfrac{2\pi}{\omega_0},
\end{equation}
where $\Phi = 1/(\tau_\mathrm{th} \omega_0)$. Relaxation time $\tau_\mathrm{th} = C_\mathrm{th} R_\mathrm{th} $ represents how quickly the system returns to thermal equilibrium. If $1 / \tau_{th} < \omega_0$, the energy loss due to heat transport must be taken into account. The basin has two channels to exchange heat with the surroundings: helium flow through the channels or thermal conduction through the substrate. The thermal conductivity through helium in the flow channels is negligible \cite{Souris2017}. The dominant limiting factor for heat conduction through the substrate is the interfacial thermal resistance (Kapitza resistance) $R_k$. The temperature dependence for the interface between liquid helium and aluminum is $R_k = 28.57 T^{-4.21}/A$ \cite{Souris2017}, where $A$ is the basin area in cm$^2$. Taking into account both basin plates, the total thermal resistance is $R_\mathrm{th} = R_k / 2$. It should be noted, however, that the Kapitza resistance is sensitive to surface preparation, and considerable scatter exists in the reported empirical values.

The total resonance full width at half maximum (FWHM) is given by the sum of \eqref{delta n} and \eqref{delta th}. Due to the end effects of the flow in the channels (i.e., the finite flow in the basin), the temperature oscillation inside the basin will not be uniform. Therefore, we use an effective basin area such that $A \to A/\lambda$, where $\lambda$ is an adjustable parameter. The only adjustable parameters are the scaling parameter $\lambda$ and the temperature correction (the same as for the frequency, i.e., parameter $b$ in Tab.~\ref{tab:parameters}). Best-fit values are listed in Tab.~\ref{tab:parameters} for the four geometries.

The resulting fit to the widths of the experimentally measured resonance curves is essentially independent of the geometry of the channel, and thus Fig.~\ref{fig:frequencies}(b) shows the average of the four fits as a single dashed line. The discrepancy between measured data and theoretical prediction is mainly in the low-temperature range; the theory underestimates the measured value by approximately a factor of two. This is likely due to uncertainty in Kapitza resistance (which is also reflected in the relatively large value of $\lambda$ in Tab.~\ref{tab:parameters}) since it strongly depends on the purity of the Al electrode and the structure of its surface.

\section{Numerical simulations}
\label{sec:app-simulations}
In order to better understand the behavior of the confined oscillatory flow, we performed two complementary sets of independent numerical simulations. First, a classical fluid dynamics calculation of confined incompressible turbulence based on the $k-\varepsilon$ model \cite{Davidson2015}; and second, a vortex filament simulation \cite{Hanninen2014} of individual quantized vortices.

\subsection{$k-\varepsilon$ simulation}
The $k-\varepsilon$ is a simplified turbulence model based on the eddy viscosity approach \cite{Davidson2015}. It is a classical continuous model that does not contain any features of superfluid dynamics (e.g., mutual friction or the discreteness of superfluid vorticity), and the obtained results, therefore, have only an illustrative character. Turbulent flows of superfluid helium with large vortex densities polarized in large structures are, however, known to behave quasi-classically \cite{Babuin2014}. 

Finite element simulation of oscillatory flow in quasi-2D Helmholtz resonators was performed for the four geometries used in the experiment (shown in Fig.~\ref{fig:devices}). The height of the cavity was set to 0.05~mm in order to maintain reasonable mesh resolution in the direction perpendicular to the plane of the device. An effective dynamical viscosity of the turbulent superfluid $^4$He $\eta_\mathrm{eff} \approx 2.9 \cdot 10^{-6}$~Pa$\cdot$s was used, which corresponds to the experimental values obtained in turbulent channel flow \cite{Babuin2014}. The boundary conditions of all walls were chosen as free-slip. At the end of the channel, the pressure was set to zero. Furthermore, to mimic the oscillatory motion of the basin plates, the normal velocity was prescribed on these plates by the function $v_{z}(x,y,t) = \pm v_0 \sin{ (\omega t)} (1 - (x^2 + y^2)/R^2)$, where $v_0$ is the amplitude, $x$ and $y$ are the in-plane coordinates, $t$ the time, $\omega$ the angular frequency, and $R$ the radius of the basin; $+$ corresponds to the bottom wall of the basin, $-$ to the upper.

\begin{figure}
    \centering
    \includegraphics{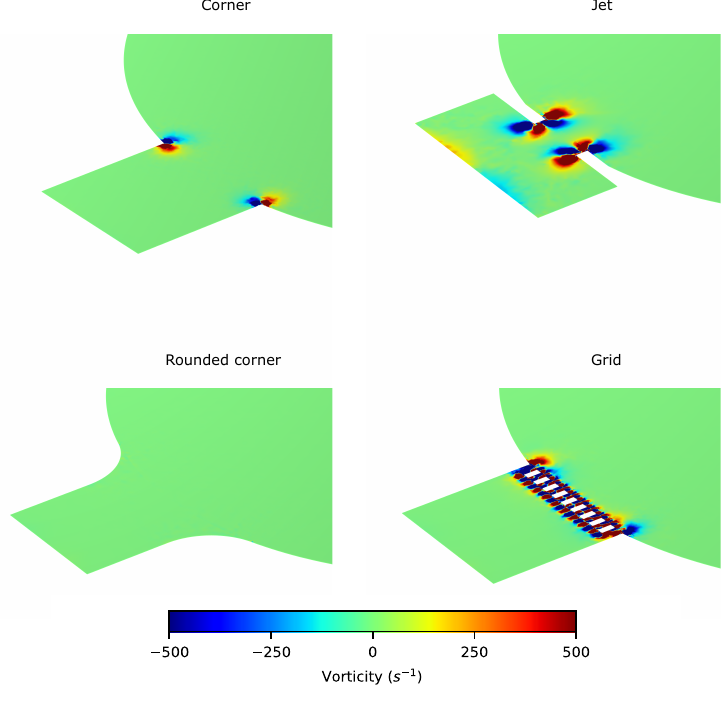}
    \caption{Vorticity of the flow averaged over one period of oscillation in quasi 2D finite element numerical simulation based on the $k - \varepsilon$ turbulence model for all four chip geometries used in the experiment. The simulation was performed using the COMSOL FEM suite.}
    \label{fig:turb_simul}
\end{figure}

To obtain time-averaged behavior, 20 periods of the oscillatory flow were simulated with velocity amplitude $v_{0} = 8 \times 10^{-4}$~m/s and the frequency $f = \omega/(2 \pi) = 2000$~Hz, for the four geometries, which correspond to the typical frequency of the Helmholtz mode and velocity in the turbulent regime. Since the experiment measures behavior averaged over many oscillations, the calculated vorticity was averaged over the last period of the calculation (when the flow was already in an oscillatory steady state). The results are shown in Fig. \ref{fig:turb_simul}. As can be seen in the cases of \emph{C}, \emph{J}, and \emph{G}, large vortical structures are observed, which are absent in the case of \emph{R}. These structures most likely correspond to the acoustic streaming patterns \cite{Duda2017}. Note that the $k-\varepsilon$ model cannot simulate the inverse cascade since it assumes isotropic dissipation \cite{Davidson2015}. However, the results show that over long time periods, the acoustic flow drives a particular mode of the spatial vorticity distribution, which is consistent with the dynamical model presented in the main text.

\subsection{Vortex filament model simulations}

The motion of individual quantized vortices in uniformly oscillating superflow and stationary normal fluid confined between two walls was numerically simulated using the vortex filament model \cite{Hanninen2014}. Quantized vortices are represented as infinitesimally thin spatial curves, which move and deform under externally applied potential superflow and the superfluid velocity field $\bv v_s(\bv r)$ induced by the vortices given by the Biot-Savart integral,
\begin{equation}
    \label{eq:vfm-biot-savart}
    \bv v_s(\bv r) = \frac{\kappa}{4\pi}\int_{L}\frac{\bv s'\bv\times(\bv r - \bv s)}{|\bv r - \bv s|^3}\mathrm{d}\xi,
\end{equation}
where $\bv s$ is a point on the vortex line, $\bv s'$ is the unit tangent vector, $\xi$ is the arc-length measured along the vortex and the integration runs through the entire system of vortices. The equation of motion of any point $\bv s$ on the vortex is
\begin{equation}
    \pdiff{\bv s}{t} = \bv v'_s + \alpha\bv s'\times (\bv v'_n - \bv v_s) - \alpha'\bv s'\times \left[\bv s'\times (\bv v'_n - \bv v_s)\right],
\end{equation}
where $\alpha$ and $\alpha'$ are temperature-dependent mutual friction coefficients \cite{Donnelly1998} and $\bv v'_s$ is the standard de-singularization of the Biot-Savart integral \eqref{eq:vfm-biot-savart} for $\bv r$ on a vortex line \cite{Hanninen2014}
\begin{equation}
    \bv v'_s(\bv r) = \frac{\kappa}{4\pi}\bv s'\times\bv s''\ln\left(\frac{2\sqrt{l_+l_-}}{e^{1/2} a}\right) + \frac{\kappa}{4\pi}\int_{L'}\frac{\bv s'\bv\times(\bv r - \bv s)}{|\bv r - \bv s|^3}\mathrm{dd}\xi,
\end{equation}
where $\bv s''$ is the local curvature vector, $l_\pm$ are the distances to forward and reverse discretization points on the vortex, and $a$ is the vortex core parameter; the integration runs over all vortices with the exception of the immediate neighborhood of $\bv r$. The vortex filament model cannot describe the nucleation of vorticity near walls; therefore, vortices are added as an initial condition. The numerical study of the experimental system with a realistic density of vortices is beyond the scope of the present work; therefore, we study the dynamics of a small number of vortices confined between two parallel walls in a doubly periodic domain in order to check the dynamical importance of three-dimensional deformations. 

The vortices are discretized to a set of connected points whose spacing is dynamically maintained between $10^{-8}$ and $1.2\times 10^{-8}$~m (simulations were checked with half the discretization length and yielded identical results). The time evolution was implemented using the 4th order Runge-Kutta stepping with fixed time step $\delta t = 10^{-10}$ s. Planar walls placed at $z=0$ and $z=D$ are considered to be uniformly pinning; a discretization point of a vortex attached to the wall is not allowed to move. However, the vortex can \emph{reconnect} with its image on the other side of the wall, which exists because of boundary conditions of no flow normal to the wall. Depinning is handled entirely by modeling pinning and reconnections, ensuring self-consistency with the rest of the model. As a result, for a sufficiently sharp angle at the contact between the vortex and the wall, the vortex can effectively slide in a manner similar to ref \cite{Nakagawa2023}.

\begin{figure}
    \centering
    \includegraphics{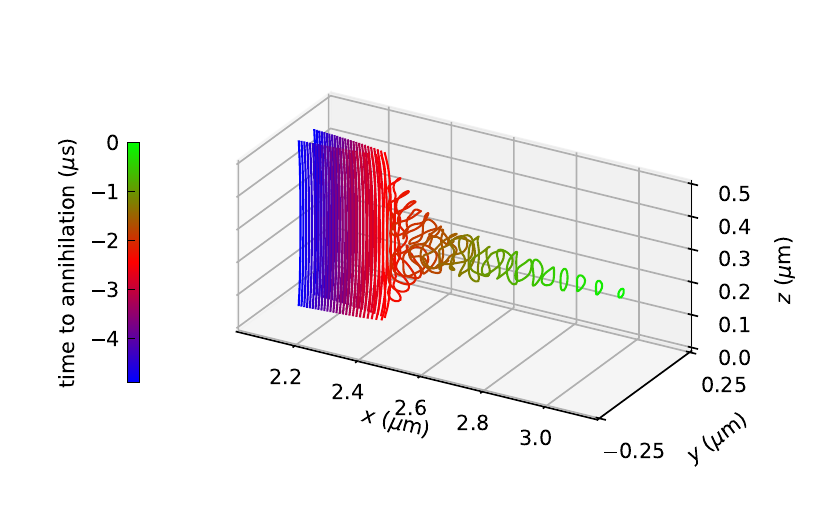}
    \caption{Typical evolution of two colliding vortices spanning the 500 nm gap between the top and bottom walls. As an initial condition, two straight vortices of opposite signs separated by 500 nm along the $y$ direction were used. The vortices are subject to a steady flow of the superfluid component with a velocity of 25 cm/s along the $x$ axis. The figure shows several snapshots leading up to the annihilation (time given by the color), with 25 ns between subsequent snapshots. The entire evolution from the initial condition to the final annihilation takes about 40 $\mu$s (the entire lead-up to the reconnection is not shown).}
    \label{fig:vortex-line-collision}
\end{figure}

\begin{figure}
    \centering
    \includegraphics{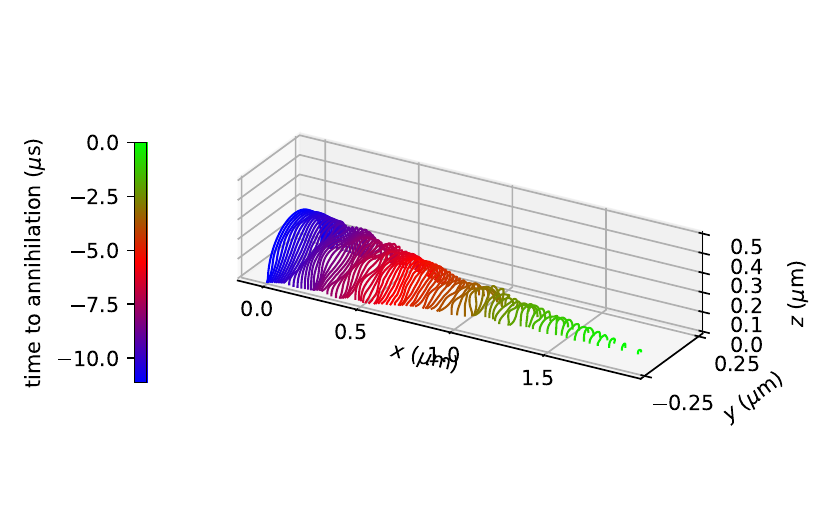}
    \caption{Annihilation of vortex half-ring with 500 nm diameter attached to the bottom wall with orientation parallel with the imposed superflow oriented along the positive x-axis direction. The initial condition was a half-circle lying in the $x=0$ plane.}
    \label{fig:ring-annihilation-plus}
\end{figure}

\begin{figure}
    \centering
    \includegraphics{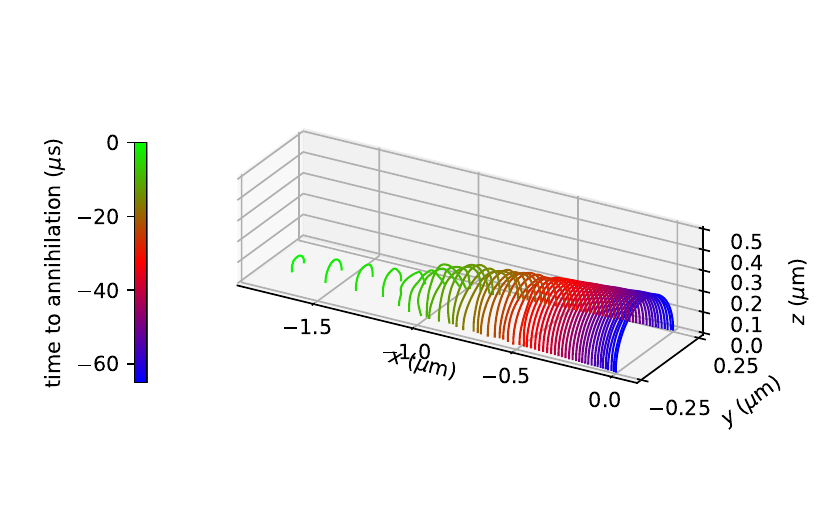}
    \caption{As in Fig.~\ref{fig:ring-annihilation-plus} with the ring orientation anti-parallel with the imposed $\bv v_s$. The vortex propagates in the negative $x$ direction because its self-induced velocity exceeds the imposed superfluid velocity.}
    \label{fig:ring-annihilation-minus}
\end{figure}

We studied the evolution of two particular scenarios with mutual friction constants corresponding to 1.45~K, $\alpha=0.061$ and $\alpha'=1.746\times 10^{-2}$ and $D = 500$~nm -- a half-loop attached to the bottom wall and a pair of reconnecting antiparallel vortex lines spanning the two walls. For all simulated cases, the imposed velocity was set to $v_s=0.25$ m/s along the x-axis and $v_n=0$. The time evolutions of the vortex configurations are shown in Fig.~\ref{fig:vortex-line-collision} for a pair of antiparallel vortices and Figs.~\ref{fig:ring-annihilation-plus}, \ref{fig:ring-annihilation-minus} for half-loops attached to the bottom wall oriented parallel or antiparallel with the imposed superfluid velocity $\bv v_s$. All initial vortex configurations studied annihilate in a few tens of microseconds with a typical flow oscillation period of 0.5~ms (see Fig.~\ref{fig:frequencies}(a)). These results suggest that the vortex configuration will remain quasi-two-dimensional, with only transient three-dimensional structures short-lived compared to the flow oscillation period.

%\bibliography{bifurcations2024}

\end{document}